\documentclass{aa}  

\usepackage{graphicx}
\usepackage{txfonts}
\usepackage{hyperref}
\usepackage{amssymb}
\usepackage{amsmath}
\usepackage{graphicx}
\usepackage{natbib}
\usepackage{rotating}
\usepackage{epsfig}
\usepackage{subfig}
\usepackage{colortbl}
\usepackage{lscape}
\usepackage{tabularx}
\usepackage{longtable}
\usepackage{caption}
\usepackage{color}

\usepackage{txfonts}

\begin{document}

   \title{Probing the innermost regions of AGN jets and their magnetic fields with \emph{RadioAstron}}
   \subtitle{II. Observations of 3C\,273 at minimum activity}

   \author{G.~Bruni\inst{1,2,3}
          \and
          J.~L.~G\'omez\inst{4}
           \and
           C.~Casadio\inst{1}      
           \and
           A.~Lobanov\inst{1,5}
          \and
          Y.~Y.~Kovalev\inst{6,1}
          \and
          K.~V.~Sokolovsky\inst{7,8,6}    
          \and
          M.~M.~Lisakov\inst{6}
          \and 
          U.~Bach\inst{1}
          \and
          A.~Marscher\inst{9}
          \and
          S.~Jorstad\inst{9,10}
          \and
          J.~M.~Anderson\inst{11}
          \and
          T.~P.~Krichbaum\inst{1}
          \and
          T.~Savolainen\inst{1,12,13}
          \and    
          L.~Vega-Garc\'ia\inst{1}
          \and
          A.~Fuentes\inst{4}      
          \and            
          J.~A.~Zensus\inst{1}  
          \and            
          A.~Alberdi\inst{4}
          \and
          S.-S.~Lee\inst{14,15}
          \and
          R.-S.~Lu\inst{1}
          \and
          M.~P\'erez-Torres\inst{4}
          \and
          E.~Ros\inst{1,16,17}
          }

   \institute{Max-Planck-Institut f\"ur Radioastronomie, Auf dem H\"{u}gel, 69, 53121  Bonn, Germany \\
              \email{gabriele.bruni@iaps.inaf.it}
   \and INAF-Istituto di Radioastronomia, via Piero Gobetti, 101, 40129 Bologna, Italy
   \and INAF-Istituto di Astrofisica e Planetologia Spaziali, via Fosso del Cavaliere, 100, 00133 Rome, Italy
   \and Instituto de Astrof\'isica de Andaluc\'ia, CSIC, Glorieta de la Astronom\'ia s/n, 18008 Granada, Spain
   \and Institut f\"{u}r Experimentalphysik, Universit\"{a}t Hamburg, Luruper Chaussee, 149, 22 761 Hamburg, Germany
   \and Astro Space Center of Lebedev Physical Institute, Profsoyuznaya 84/32, 117997 Moscow, Russia
   \and IAASARS, National Observatory of Athens, Vas. Pavlou \& I. Metaxa, 15236 Penteli, Greece
   \and Sternberg Astronomical Institute, Moscow State University, Universitetskii pr. 13, 119992 Moscow, Russia
   \and Institute for Astrophysical Research, Boston University, 725 Commonwealth Avenue, Boston, MA 02215
   \and Astronomical Institute, St. Petersburg State University, Universitetskij Pr. 28, Petrodvorets, 198504 St. Petersburg, Russia
   \and Deutsches GeoForschungsZentrum GFZ, Telegrafenberg, 14473 Potsdam, Germany
   \and Aalto University Mets\"ahovi Radio Observatory, Mets\"ahovintie 114,02540 Kylm\"al\"a, Finland
   \and Aalto University Department of Electronics and Nanoengineering, PL 15500, FI-00076 Aalto, Finland
   \and Korea Astronomy and Space Science Institute, 776 Daedeok-daero, Yuseong-gu, 34055 Daejeon, Korea
   \and Korea University of Science and Technology, 217 Gajeong-ro, Yuseong-gu, Daejeon 34113, Korea
   \and Observatori Astron\'omic, Universitat de Val\`encia, Parc Cient\'{\i}fic, C.\ Catedr\'atico Jos\'e Beltr\'an 2, E-46980 Paterna, Val\`encia, Spain
   \and Departament d'Astronomia i Astrof\'isica, Universitat de Val\`encia, 46100 Burjassot, Val\`encia, Spain
   }

   \date{}

 
  \abstract
   {\emph{RadioAstron} is a 10 m orbiting radio telescope mounted on the \emph{Spektr-R} satellite, launched in 2011, performing Space Very Long Baseline Interferometry (SVLBI) observations supported by a global ground array of radio telescopes. With an apogee of $\sim$350\,000 km, it is offering for the first time the possibility to perform $\mu$as-resolution imaging in the cm-band.}
   {The \emph{RadioAstron} Active Galactic Nuclei (AGN) polarization Key Science Project (KSP) aims at exploiting the unprecedented angular resolution provided by \emph{RadioAstron} to study jet launching/collimation and magnetic-field configuration in AGN jets. The targets of our KSP are some of the most powerful blazars in the sky.}
   {We present observations at 22 GHz of 3C\,273, performed in 2014, designed to reach a maximum baseline of approximately nine Earth diameters. Reaching an angular resolution of 0.3 mas, we study a particularly low-activity state of the source, and estimate the nuclear region brightness temperature, comparing with the extreme one detected one year before during the \emph{RadioAstron} early science period. We also make use of the VLBA-BU-BLAZAR survey data, at 43 GHz, to study the kinematics of the jet in a $\sim$1.5-year time window.}
   {We find that the nuclear brightness temperature is two orders of magnitude lower than the exceptionally high value detected in 2013 with \emph{RadioAstron} at the same frequency (1.4$\times10^{13}$ K, source-frame), and even one order of magnitude lower than the equipartition value. The kinematics analysis at 43 GHz shows that a new component was ejected $\sim$2 months after the 2013 epoch, visible also in our 22 GHz map presented here. Consequently this was located upstream of the core during the brightness temperature peak. \emph{Fermi}-LAT observations for the period 2010-2014 do not show any $\gamma$-ray flare in conjunction with the passage of the new component by the core at 43 GHz.}
   {These observations confirm that the previously detected extreme brightness temperature in 3C~273, exceeding the inverse Compton limit, is a short-lived phenomenon caused by a temporary departure from equipartition. Thus, the availability of interferometric baselines capable of providing $\mu$as angular resolution does not systematically imply measured brightness temperatures over the known physical limits for astrophysical sources.}


   \maketitle
%

\section{Introduction}

More than 50 years have passed since Maarten Schmidt's discovery of the first quasi-stellar object: 3C\,273 (\citealt{Schmidt}). Later classified as Flat Spectrum Radio Quasar (FSRQ), and thus belonging to the blazar population, it presents a strong and structured jet in the mm/cm band, with a viewing angle of 5-11 degrees (\citealt{Liu}), that dominates the emission up to the infrared band. 
\cite{Lobanov01} presented the first Space Very Long Baseline Interferometry (SVLBI) study of 3C\,273, using a ground array composed of ten Very Long Baseline Array (VLBA) antennas, plus the Effelsberg 100-m single dish, and the 8 m orbiting dish \emph{VSOP} on-board the \emph{HALCA} satellite (VLBI Space Observatory Programme, \citealt{Hirabayashi}). Achieving an angular resolution of 2.1$\times$0.5 mas at 5 GHz, they could identify a double-helical structure, consistent with a Kelvin-Helmholtz instability in the jet stream.  

\emph{RadioAstron} (RA hereafter) is the first mission for SVLBI after the \emph{VSOP} era (1990s). Launched in July 2011, it is led by Astro Space Center (ASC, Moscow, Russia) and features a 10 m dish on board the satellite \emph{Spektr-R} (\citealt{Kardashev}). Supported by a global ground array of radio telescopes, it can reach space baselines as long as 350\,000 km.
The RA Key Science Project (KSP) on Active Galactic Nuclei (AGN) polarization aims at developing, commissioning, and exploiting the unprecedented high-angular-resolution polarization capabilities of RA to probe the innermost regions of AGN jets and their magnetic fields. The KSP observations have been performed/scheduled over the first four observing periods of RA (AO-1/2/3/4, 2013-2017). Observations have so far targeted a sample of the most active and highly polarized AGNs in the sky (BL\,Lac, 3C\,273, 3C\,279, OJ\,287, 0716+714, 3C\,345, 3C\,454.3, CTA\,102). The first polarimetric test observations of 0642+449 were performed on March 9-10, 2013 (early science period) at 1.6 GHz (L-band), with participation of the European VLBI Network (EVN) including the Kvazar network and the telescopes in Evpatoria and Green Bank. The correlated signal between the ground telescopes and RA was detected on projected baselines of up to six Earth diameters in length, achieving a resolution of 1 mas at the respective fringe-spacing of $\sim$400 M$\lambda$.
The RA instrumental polarization was found to remain stable, consistent across the different IFs throughout the experiment, and to be within 9\%, demonstrating the excellent polarization capabilities of RA (\citealt{Lobanov}). 


The first 22 GHz observations of our KSP were performed in November 2013, targeting BL Lac in an array that included 15 radio telescopes on the ground (\citealt{gomez}). The instrumental polarization of the space radio telescope was found to be smaller than 9\%, confirming the polarimetric capabilities of RA at 22 GHz. Correlated visibilities between the ground antennas and the space radio telescope were found extending up to a projected baseline distance of 7.9 Earth diameters ($D_\mathrm{E}$ hereafter), yielding a maximum angular resolution of 21 $\mu$as,  the highest achieved to date (\citealt{gomez}). At these angular scales, not probed before in a blazar jet, evidence for emission upstream of the VLBI core was found, being interpreted as corresponding to a recollimation shock at about 40~$\mu$as from the jet apex, in a pattern that includes also two more recollimation shocks at 100 and 250 $\mu$as. Linearly polarized emission was clearly detected in two components within the innermost 0.5~mas from the core, as well as in the downstream jet. Combination of RA 22 GHz and simultaneous ground-based 15 and 43 GHz images allowed for the mapping of the Faraday rotation in the jet of BL~Lac, revealing a gradient in rotation measure and polarization vectors as a function of position angle with respect to the centroid of the core, in agreement with the existence of a large-scale helical magnetic field threading the jet in BL Lac. The intrinsic de-boosted brightness temperature of the unresolved core was found to be in excess of $3\!\times\!10^{12}$~K (\citealt{gomez}), exceeding both the equipartition value ($5\!\times\!10^{10}$~K, \citealt{Readhead}) and the inverse Compton cooling limit $10^{11.5}$~K (\citealt{Kellermann}).

Recently, \cite{Kovalev} found an extremely high brightness temperature for 3C\,273, using non-imaging observations at 22 GHz from the AGN-survey KSP. In the period Dec. 2012 - Feb. 2013, values in excess of 10$^{13}$ K were measured, thus again larger than the theoretical limit imposed by inverse Compton cooling. The maximum baseline reached by the experiment was 171\,000 km. RA can probe structures at a resolution never achieved before, finding brightness temperature values that challenge our understanding of non-thermal emission in AGN.
 
This work is the first of two papers presenting observations of 3C\,273 carried out in the context of the RA AGN polarization KSP. Here we focus on observations at 22 GHz, and the comparison with the results from \cite{Kovalev}, while 1.6 GHz observations will be presented in a future paper.

For a flat Universe with $\Omega_{\rm{m}}$= 0.3, $\Omega_{\Lambda}$= 0.7, and $H_{0}$ = 70 km s$^{-1}$ Mpc$^{-1}$ \citep{Planck}, 1 mas corresponds to {2.729} pc at the redshift of 3C\,273 ($z$=0.158).

\section{Observations and data processing}

The source was targeted twice during AO-1 in the framework of the AGN polarization KSP, with a global ground array plus RA. 
In the following, details about data correlation and calibration are given.

\subsection*{22 GHz observations}
\begin{figure}
\includegraphics[width=9cm]{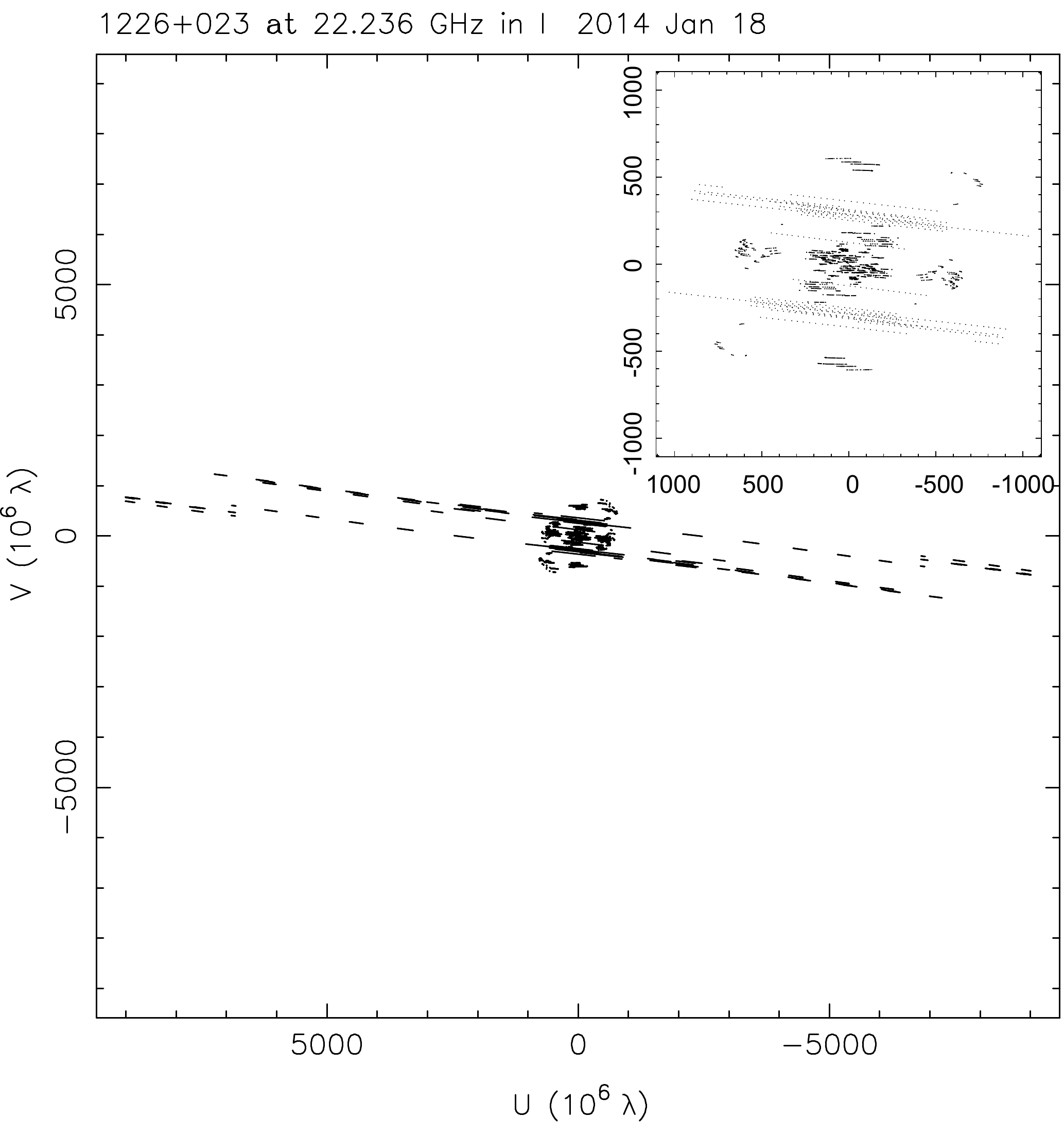}
\caption{UV coverage for 3C\,273 at 22 GHz. The ground global array, covering the Earth diameter, is visible in the middle, while wings result from RA baselines. In the inset, a zoom on the part giving space-fringes is shown.}
\label{UV-K}
\end{figure}

A global ground array of 22 antennas was used to perform observations, including VLBA (Sc, Hn, Nl, Fd, La, Kp, Pt, Ov, Br, Mk), EVN (Hh, Mc, Nt, Tr, Jb, Ef, Ys), Long Baseline Array (-LBA- At, Mp, Ho, Cd), and two Kvazar antennas (Sv, Zc), plus Kalyazin (managed by ASC, Russia), and Green Bank (NRAO, USA). The observations took place on January 18-19, 2014, for a total of 16.8 hours, and at three different frequencies: 15 GHz, 22 GHz, and 43 GHz. RA was involved only for the 22 GHz part, while for the other bands only the VLBA was used. Both the Green Bank and Pushchino tracking station took part in the experiment. A RA-compatible total bandwidth of 32 MHz, split into two 16-MHz IFs, was used. RA was scheduled to observe three consecutive 9.5 minute scans every 1.25 hours, to allow for antenna cooling. The maximum projected space-baseline reached during the experiment was $\sim$9 G$\lambda$ ($\sim$117\,000 km, $\sim$9 Earth diameters). Fig. \ref{UV-K} reports the scheduled UV coverage, as well as the one giving fringes with RA in the inset. Effelsberg was also used in single-dish mode, to perform total flux density and polarimetry measurements of the target and calibrators from January 10 to 28, 2014, at 5, 10, and 22 GHz.

\begin{figure*}
\begin{center}
\includegraphics[width=12cm]{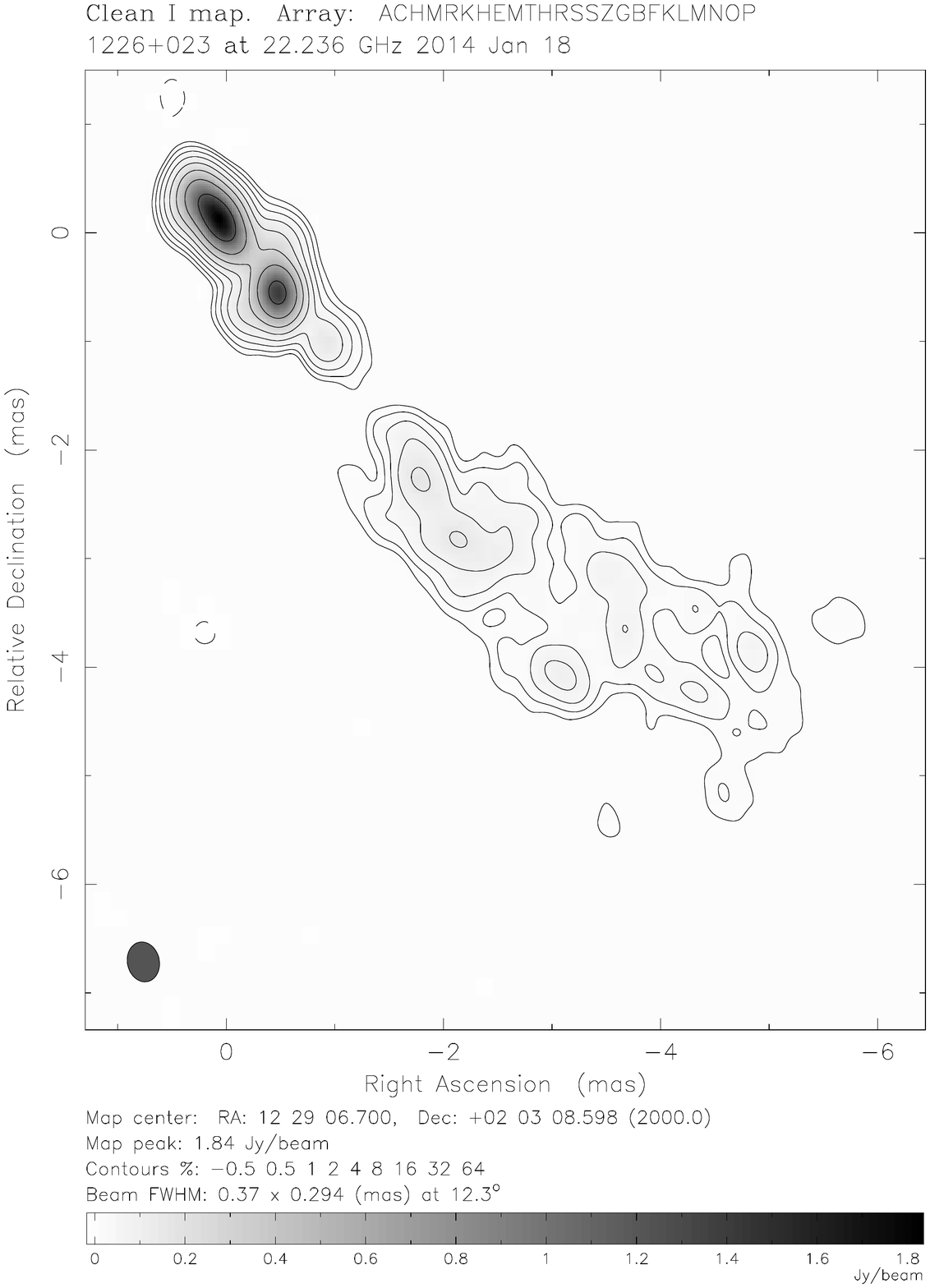}
\caption{Map of 3C\,273 at 22 GHz obtained with the global ground array plus RA, using uniform weighting. The lowest contour significance is 5-sigma.}
\label{K}
\end{center}
\end{figure*}

\subsection*{Data processing and reduction}

Correlation was performed at the MPIfR-Bonn correlator, using the \emph{ra} version of DiFX for space-VLBI (\citealt{Deller,Bruni}). Fringe searching for RA was attempted for every scan, covering the entire duration of the experiment, in order to optimize the centering of the correlation window. A suitable reference antenna was used depending on the available arrays (ATCA for the LBA, Effelsberg for the EVN, Green Bank for VLBA). Fringes with RA with a signal-to-noise ratio (SNR) above 10 were found only for the last hour of the experiment, when the spacecraft was near to perigee. In fact, 3C\,273 had a flux density historical minimum across the observations (see Sect. \ref{sec:OVRO}). This resulted in a maximum baseline for space-VLBI detections of $\sim$1 G$\lambda$. 

Data were reduced using the Astronomical Image Processing System ({\tt AIPS}\footnote{\url{http://www.aips.nrao.edu/index.shtml}}) software for dataset calibration and fringe searching, and the Differential Mapping software ({\tt Difmap}\footnote{\url{https://science.nrao.edu/facilities/vlba/docs/manuals/oss2013a/post-processing-software/difmap}}) for the imaging part. Following the recipe developed for RA imaging datasets, the ground array was phased first in {\tt AIPS}, and only after applying the solutions found in this first step. Fringes between RA and the global ground array were searched for, allowing baseline stacking and exhaustive baseline searching in the {\tt FRING} task (\citealt{gomez}). Thanks to this method, ground-space fringes were found in two more scans with respect to the correlation-stage, but not significantly extending the ground-space baselines. For all other RA scans, no fringes were found.
Subsequently, data were imported into {\tt Difmap} for imaging. We flagged all baselines to RA corresponding to time ranges with non-valid {\tt FRING} solutions. Attempts were made to use clean components from a source map from ground array as the model for fringe searching in {\tt AIPS}, also merging IFs and polarizations to improve sensitivity, but no further signal was found. A well-detected polarized signal was found only on LBA baselines, so we decided to drop polarimetry analysis from this work, not adding significant information.


\begin{table*}
\centering
\caption{Flux density, size, observed and intrinsic $T_\mathrm{b}$ for the components used to modelfit the core region of the 22 GHz RA map.}
 \begin{tabular}{cccccc}
  \hline
Comp.   &  Flux density         &  Size         & $T_\mathrm{b,obs}$                    & $T_\mathrm{b,int}$              & Dist. from A  \\
                &  [Jy]         &  [mas]        & [$10^{10}$ K]                 & [$10^{10}$ K]   & [mas]                 \\
  \hline
A &  0.74 $\pm$ 0.02            &   0.170 $\pm$ 0.008           &   5.89 $\pm$ 0.56              &       0.52 $\pm$ 0.05  &      --      \\
B &  2.66 $\pm$ 0.08            &   0.260 $\pm$ 0.013           &   9.04 $\pm$ 0.91              &       0.81 $\pm$ 0.08  &      0.37 $\pm$ 0.13 \\
C &  1.94 $\pm$ 0.06            &   0.153 $\pm$ 0.007           &  19.0 $\pm$ 1.7~~   &       1.70 $\pm$ 0.16  &      1.19 $\pm$ 0.08 \\
  \hline
 \end{tabular}
 \label{model}
\end{table*}

\section{Results}

\begin{figure}[]
\centering
\includegraphics[width=0.5\textwidth]{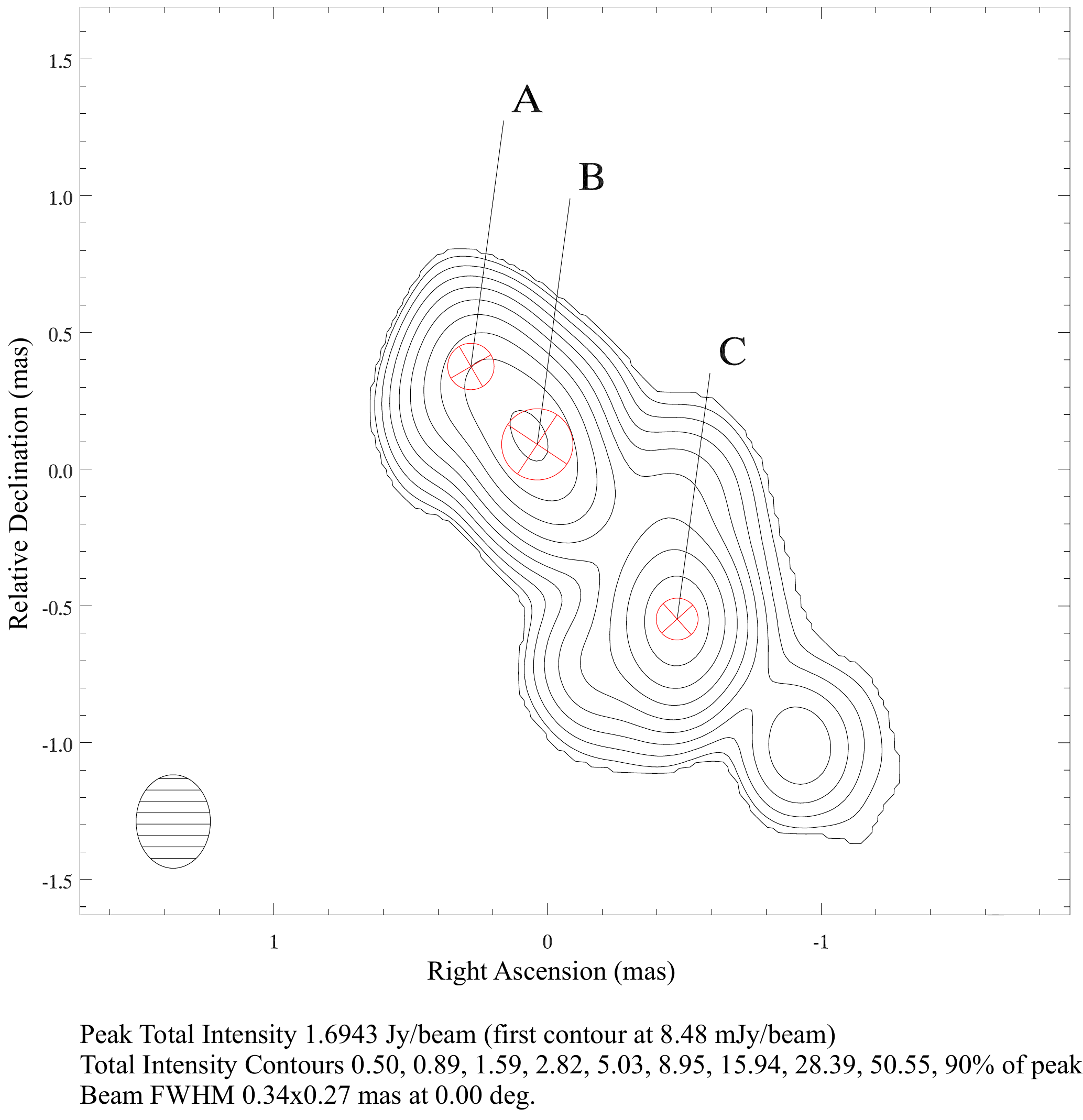}
\caption{Zoom on the central region of 3C273 at 22 GHz (RA): image obtained using three modelfit Gaussian components for the $T_\mathrm{b}$ estimate (overplotted).}
\label{corezoom}
\end{figure}

The obtained map at 22 GHz is shown in Fig. \ref{K}. The angular resolution, using uniform weighting, is 0.37$\times$0.29 mas at a position angle of 12.3 degrees.

\subsection{Brightness temperature estimate}

The emission from the central region was modeled in {\tt Difmap} to extract the flux density and linear size of the fitted Gaussian components (see Fig. \ref{corezoom}). The extracted values are given in Table \ref{model}. The observed brightness temperature ($T_\mathrm{b,obs}$) at a certain wavelength $\lambda$ (cm) can be estimated from the measured flux density $S$ (Jy) and size $\theta$ (mas) of a circular Gaussian component, using the following formula (e.g. \citealt{gomez}):
\begin{equation}
T_\mathrm{b,obs}=1.36\times10^9\frac{S\lambda^2}{\theta^2}~~[\mathrm{K}],
\end{equation}
and it is related to the intrinsic brightness temperature ($T_\mathrm{b,int}$) of the source via the following formula:
\begin{equation}
T_\mathrm{b,int}=\frac{T_\mathrm{b,obs}(1+z)}{\delta}~~[\mathrm{K}],
\end{equation}
where $\delta$ is the Doppler factor of the source (derived from the jet speed), estimated from variability or kinematics analysis, and $z$ the redshift. For 3C\,273, $z$=0.158, while $\delta$ has been estimated to be $\le$13 by \cite{Jorstad}. For our calculation, we consider the maximum value ($\delta$=13), resulting in a minimum value for our $T_\mathrm{b,int}$ estimate. The obtained values are shown in Table \ref{model}, the uncertainties for size, flux density, and position of the components have been calculated following \cite{Jorstad}.
These values are well below the one imposed by the inverse Compton cooling limit ($10^{11.5}$ K), and for the core (A) even one order of magnitude lower than the typical value in equipartition conditions ($5\times10^{10}$ K). This indicates that the source is in an exceptionally low-activity state.

We want to stress that this estimate is not limited by baseline length - and thus angular resolution - since the maximum space-baseline of the experiment was $\sim$9 G$\lambda$, that is, comparable (even larger) to the one that led to the extreme brightness temperature measured at the same frequency by \cite{Kovalev} (7.6 G$\lambda$). The fact that fringes are not detected at more than $\sim$1 G$\lambda$ on RA baselines is then caused by the effective low flux density from the source in such a low-activity state.







\begin{figure}[]
\centering
\includegraphics[width=0.5\textwidth]{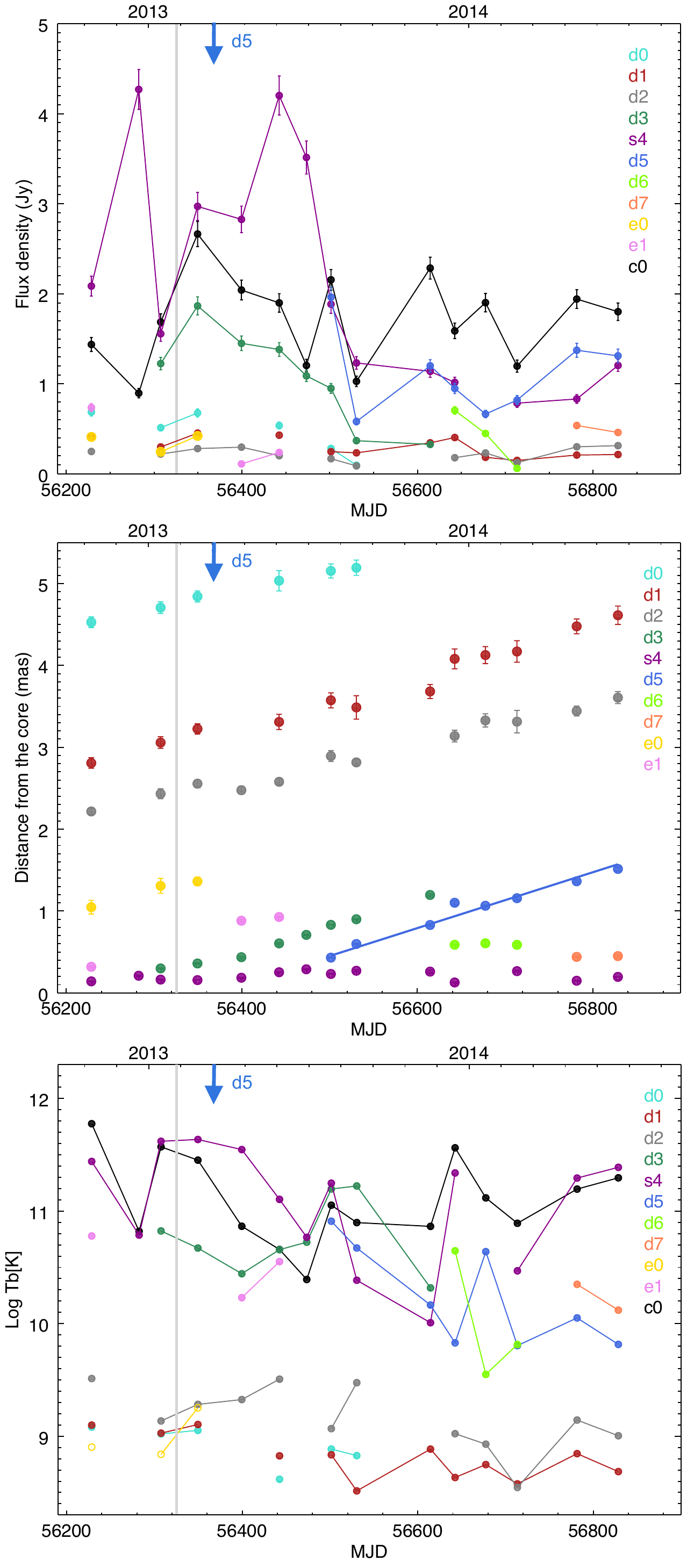}
\caption{Results from the 43 GHz VLBA-BU-BLAZAR data from 2012-2014, for each modelfit component. The gray vertical line indicates the \cite{Kovalev} epoch and the arrow indicates the ejection epoch for component \emph{d5}. Top panel: light curves; middle panel: distance from the core versus time; d5 is the radio component ejected after the \cite{Kovalev} epoch; bottom panel: observed brightness temperature versus time.}
\label{BU_combined}
\end{figure}

\begin{figure}[]
\centering
\includegraphics[width=0.5\textwidth]{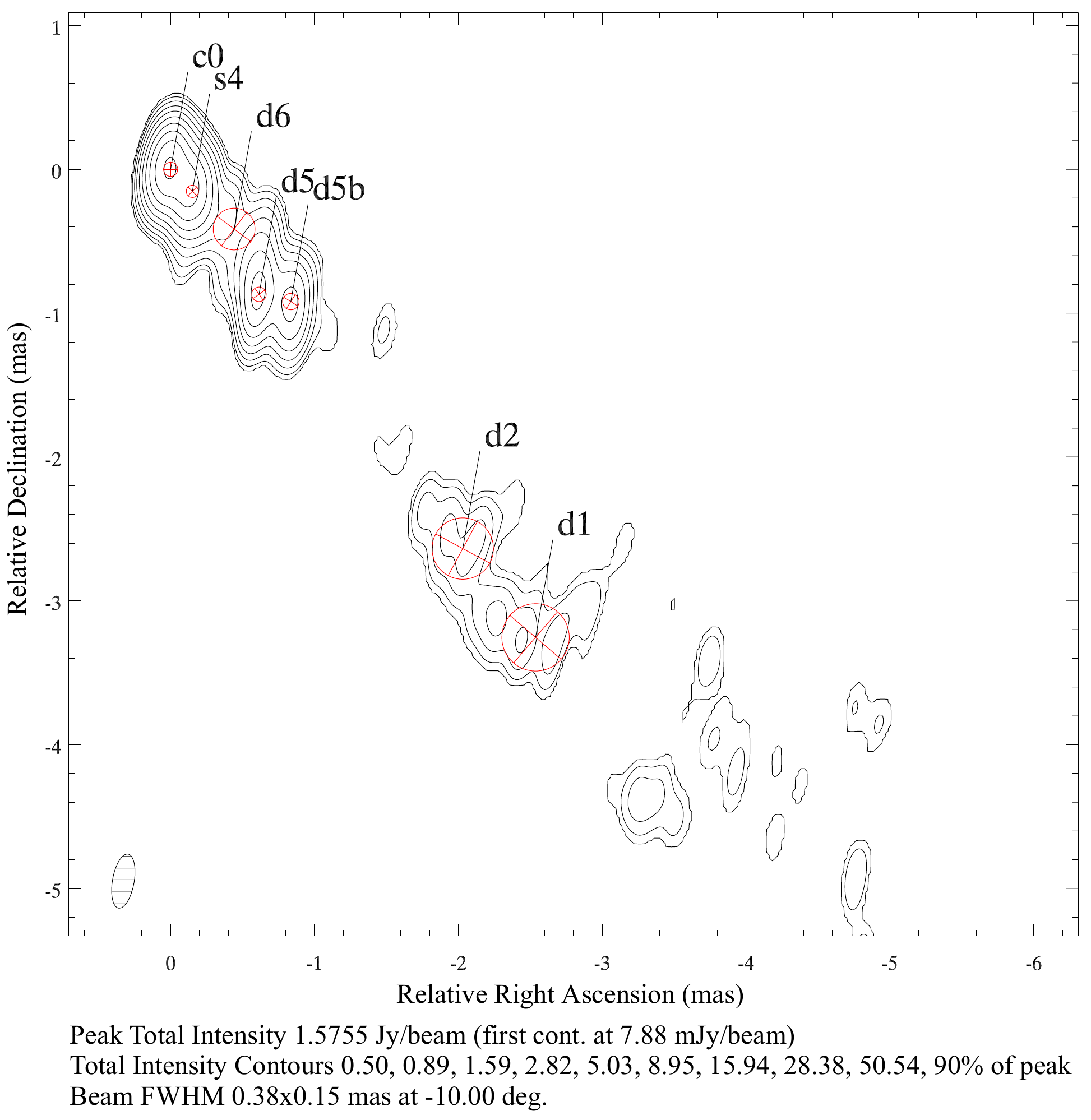}
\caption{VLBA-BU-BLAZAR total intensity image on 19th January 2014, obtained with the modelfit Gaussian components (overplotted). The angular resolution is 0.38$\times$0.15 mas.}
\label{43image}
\end{figure}

\begin{table*}
\caption{Quantities from the modelfit analysis of VLBA-BU-BLAZAR data: epoch, flux density at 43 GHz, distance from the core, position angle, major axis, and both observed and instrinsic $T_\mathrm{b}$ of the circular Gaussian components \emph{c0}, \emph{s4}, and \emph{d5} are given.}
\label{modelfit}
\begin{center}
\scalebox{0.9}{
\begin{tabular}{ l c c c c c c c }
\hline
Epoch   & Epoch         & Flux density  & Dist. from \emph{c0}  & Pos. Angle         &   Size                & $T_\mathrm{b,obs}$    & $T_\mathrm{b,int}$         \\ [0.07cm]
[year]  & [MJD]         & [mJy]         & [mas]                 & [$^{\circ}$]         &  [$\mu$as]    & [$10^{10}$ K]                 & [$10^{10}$ K]         \\ [0.07cm]
\hline
&&&& Component \emph{c0} &&\\
\hline
2012.82 & 56228.5 & 1437 $\pm$ 71               & -- & -- & 40 $\pm$ 2                   & 59.8 $\pm$ 6.0~~              &       5.33 $\pm$ 0.53 \\ [0.07cm] 
2012.97 & 56282.5 & ~~896 $\pm$ 44              & -- & -- & 95 $\pm$ 5                   &  6.62 $\pm$ 0.70              &       0.59 $\pm$ 0.06 \\ [0.07cm] 
2013.04 & 56307.5 & 1686 $\pm$ 84               & -- & -- & 55 $\pm$ 3                   & 37.1 $\pm$ 4.0~~              &       3.31 $\pm$ 0.36 \\ [0.07cm] 
2013.15 & 56349.5 & ~~2663 $\pm$ 133            & -- & -- & 79 $\pm$ 4                   & 28.4 $\pm$ 2.9~~              &       2.53 $\pm$ 0.26 \\ [0.07cm] 
2013.29 & 56399.5 & ~~2041 $\pm$ 102            & -- & -- & 136 $\pm$ 7~~         &  7.35 $\pm$ 0.76              &       0.66 $\pm$ 0.07 \\ [0.07cm] 
2013.41 & 56442.5 & 1899 $\pm$ 94               & -- & -- & 167 $\pm$ 8~~         &  4.54 $\pm$ 0.44              &       0.40 $\pm$ 0.04 \\ [0.07cm] 
2013.49 & 56473.5 & 1204 $\pm$ 60               & -- & -- & 180 $\pm$ 9~~         &  2.48 $\pm$ 0.25              &       0.22 $\pm$ 0.02 \\ [0.07cm] 
2013.57 & 56501.5 & ~~2155 $\pm$ 107            & -- & -- & 113 $\pm$ 6~~         & 11.2 $\pm$ 1.2~~              &       1.00 $\pm$ 0.11 \\ [0.07cm] 
2013.65 & 56530.5 & 1028 $\pm$ 51               &  -- & -- & 93 $\pm$ 5                  &  7.92 $\pm$ 0.85              &       0.71 $\pm$ 0.08 \\ [0.07cm] 
2013.88 & 56614.5 & ~~2285 $\pm$ 114            & -- & -- & 144 $\pm$ 7~~         &  7.34 $\pm$ 0.71              &       0.65 $\pm$ 0.06 \\ [0.07cm] 
2013.96 & 56642.5 & 1587 $\pm$ 79               & -- & -- & 54 $\pm$ 3                   & 36.3 $\pm$ 4.0~~              &       3.23 $\pm$ 0.36 \\ [0.07cm] 
2014.05 & 56677.5 & 1901 $\pm$ 95               & -- & -- & 98 $\pm$ 5                   & 13.2 $\pm$ 1.3~~              &       1.17 $\pm$ 0.12 \\ [0.07cm] 
2014.15 & 56713.5 & 1197 $\pm$ 59               & -- & -- & 101 $\pm$ 5~~         &  7.82 $\pm$ 0.77              &       0.70 $\pm$ 0.07 \\ [0.07cm] 
2014.34 & 56781.5 & 1941 $\pm$ 97               & -- & -- & 91 $\pm$ 5                   & 15.6 $\pm$ 1.7~~              &       1.39 $\pm$ 0.15 \\ [0.07cm] 
2014.47 & 56828.5 & 1801 $\pm$ 90               & -- & -- & 78 $\pm$ 4                   & 19.7 $\pm$ 2.0~~              &       1.76 $\pm$ 0.18  \\ [0.07cm] 
\hline
&&&& Component \emph{s4} &&\\
\hline
2012.82 & 56228.5 & 2085 $\pm$ 104      & 0.14 $\pm$ 0.01       & $-$135.9 $\pm$ 5.8       & 71 $\pm$ 4            & 27.6 $\pm$ 3.1                &       2.46 $\pm$ 0.28 \\ [0.07cm] 
2012.97 & 56282.5 & 4269 $\pm$ 213      & 0.21 $\pm$ 0.01       & $-$154.9 $\pm$ 3.6       & 216 $\pm$ 11          &  ~~6.10 $\pm$ 0.62    &       0.54 $\pm$ 0.06 \\ [0.07cm] 
2013.04 & 56307.5 & 1556 $\pm$ 77~~     & 0.16 $\pm$ 0.01       & $-$141.1 $\pm$ 5.0       & 50 $\pm$ 2            & 41.5 $\pm$ 3.3                &       3.69 $\pm$ 0.30  \\ [0.07cm] 
2013.15 & 56349.5 & 2970 $\pm$ 148      & 0.16 $\pm$ 0.01       & $-$138.2 $\pm$ 5.2       & 68 $\pm$ 3            & 42.8 $\pm$ 3.8                &       3.81 $\pm$ 0.34 \\ [0.07cm] 
2013.29 & 56399.5 & 2826 $\pm$ 141      & 0.19 $\pm$ 0.01       & $-$141.0 $\pm$ 4.4       & 73 $\pm$ 4            & 35.3 $\pm$ 3.9                &       3.15 $\pm$ 0.35 \\ [0.07cm] 
2013.41 & 56442.5 & 4201 $\pm$ 210      & 0.25 $\pm$ 0.01       & $-$137.0 $\pm$ 3.2       & 148 $\pm$ 7~~         & 12.8 $\pm$ 1.2                &       1.14 $\pm$ 0.11 \\ [0.07cm] 
2013.49 & 56473.5 & 3514 $\pm$ 175      & 0.29 $\pm$ 0.01       & $-$137.4 $\pm$ 2.8       & 200 $\pm$ 10          &  ~~5.85 $\pm$ 0.59    &       0.52 $\pm$ 0.05 \\ [0.07cm] 
2013.57 & 56501.5 & 1884 $\pm$ 94~~     & 0.23 $\pm$ 0.01       & $-$135.4 $\pm$ 3.5       & 84 $\pm$ 4            & 17.8 $\pm$ 1.7                &       1.58 $\pm$ 0.15 \\ [0.07cm] 
2013.65 & 56530.5 & 1232 $\pm$ 61~~     & 0.27 $\pm$ 0.01       & $-$129.0 $\pm$ 3.0       & 184 $\pm$ 9~~         &  ~~2.42 $\pm$ 0.24    &       0.22 $\pm$ 0.02 \\ [0.07cm] 
2013.88 & 56614.5 & 1139 $\pm$ 56~~     & 0.26 $\pm$ 0.01       & $-$133.8 $\pm$ 3.1       & 273 $\pm$ 14          &  ~~1.02 $\pm$ 0.11    &       0.09 $\pm$ 0.01 \\ [0.07cm] 
2013.96 & 56642.5 & 1016 $\pm$ 50~~     & 0.13 $\pm$ 0.01       & $-$130.0 $\pm$ 6.4       & 56 $\pm$ 3            & 21.6 $\pm$ 2.3                &       1.92 $\pm$ 0.21 \\ [0.07cm] 
2014.05 & 56677.5 & 1202 $\pm$ 60~~     & 0.22 $\pm$ 0.01       & $-$135.3 $\pm$ 3.8       & 85 $\pm$ 4            & 11.1 $\pm$ 1.0                &       0.99 $\pm$ 0.09 \\ [0.07cm] 
2014.15 & 56713.5 & 785 $\pm$ 39                & 0.27 $\pm$ 0.01       & $-$135.3 $\pm$ 3.1      & 133 $\pm$ 7~~         &  ~~2.96 $\pm$ 0.31    &       0.26 $\pm$ 0.03 \\ [0.07cm] 
2014.34 & 56781.5 & 832 $\pm$ 41                & 0.15 $\pm$ 0.01       & $-$132.4 $\pm$ 5.5      & 53 $\pm$ 3            & 19.7 $\pm$ 2.2                &       1.76 $\pm$ 0.20  \\ [0.07cm] 
2014.47 & 56828.5 & 1204 $\pm$ 60~~     & 0.20 $\pm$ 0.01       & $-$130.3 $\pm$ 4.1       & 57 $\pm$ 3            & 24.7 $\pm$ 2.6                &       2.20 $\pm$ 0.23 \\ [0.07cm] 
\hline
&&&& Component \emph{d5} &&\\
\hline
2013.57 & 56501.5 & 1962 $\pm$ 98~~     & 0.43 $\pm$ 0.01       & $-$133.8 $\pm$ 1.9       & 127 $\pm$ 6~~         & 8.11 $\pm$ 0.77       &       0.72 $\pm$ 0.07 \\ [0.07cm] 
2013.65 & 56530.5 & 581 $\pm$ 29                & 0.60 $\pm$ 0.01       & $-$140.4 $\pm$ 1.4      & 91 $\pm$ 5            & 4.68 $\pm$ 0.52       &       0.42 $\pm$ 0.05 \\ [0.07cm] 
2013.88 & 56614.5 & 1200 $\pm$ 60~~     & 0.83 $\pm$ 0.01       & $-$142.0 $\pm$ 1.0       & 234 $\pm$ 12          & 1.46 $\pm$ 0.15       &       0.13 $\pm$ 0.01 \\ [0.07cm] 
2013.96 & 56642.5 & 948 $\pm$ 47                & 1.10 $\pm$ 0.01       & $-$141.4 $\pm$ 0.7      & 306 $\pm$ 15          & 0.67 $\pm$ 0.07       &       0.06 $\pm$ 0.01 \\ [0.07cm] 
2014.05 & 56677.5 & 664 $\pm$ 33                & 1.06 $\pm$ 0.01       & $-$144.7 $\pm$ 0.8      & 101 $\pm$ 5~~         & 4.34 $\pm$ 0.43       &       0.39 $\pm$ 0.04 \\ [0.07cm] 
2014.15 & 56713.5 & 820 $\pm$ 41                & 1.16 $\pm$ 0.01       & $-$141.5 $\pm$ 0.7      & 293 $\pm$ 15          & 0.64 $\pm$ 0.07       &       0.06 $\pm$ 0.01 \\ [0.07cm] 
2014.34 & 56781.5 & 1373 $\pm$ 68~~     & 1.36 $\pm$ 0.01       & $-$140.1 $\pm$ 0.6       & 285 $\pm$ 14          & 1.13 $\pm$ 0.11       &       0.10 $\pm$ 0.01 \\ [0.07cm] 
2014.47 & 56828.5 & 1311 $\pm$ 65~~     & 1.51 $\pm$ 0.01       & $-$139.7 $\pm$ 0.5       & 366 $\pm$ 18          & 0.65 $\pm$ 0.06       &       0.06 $\pm$ 0.01 \\ [0.07cm]
\hline
\end{tabular}}
\end{center}
\end{table*}

\subsection{Kinematics analysis from VLBA-BU-BLAZAR data}

We also performed the analysis of VLBA data at 43 GHz taken from the VLBA-BU-BLAZAR program\footnote{\url{http://www.bu.edu/blazars/research.html}} in the observing period between October 2012 and June 2014. To compare epochs among them, the images have been convolved with a common mean beam of 0.38$\times$0.15 mas at $\mathrm{-10^{\circ}}$, corresponding to an average spatial scale comparable with the RA map. The minor beam axis is half the one in the RA map, though. The data reduction was performed with a combination of {\tt AIPS} and {\tt Difmap}, as described in \cite{Jorstad}. 

The analysis of the jet kinematics and flux density variability was carried out by fitting a series of circular Gaussian components which best model the brightness distribution of the source. Identification of components across epochs was performed assuming a smooth variation in flux density and proper motion of the model fitted components. We identified superluminal as well as stationary features. The core, identified with component \emph{c0}, is considered stationary among epochs. Another stationary component, labeled as \emph{s4}, was found very close to the core. From the light curves of model-fit components in Fig~\ref{BU_combined} (upper panel) we discovered that 3C~273 underwent a radio flare starting around the beginning of 2013, when the core (in black) increased its flux density and, subsequently, a new superluminal radio component (\emph{d5}) emerged from the core region. We inferred the time of ejection of this new component, 2013.20$\pm$0.01, that is, as expected, just after the increase of the core flux density (see Fig.~\ref{BU_combined}, middle panel). We can also notice that after having increased its brightness, the core displayed a decrease, while the stationary component \emph{s4} reached a peak and only later on is it possible to distinguish component \emph{d5} in the images. This leads us to associate component \emph{s4} with a possible recollimation shock in the core region, that becomes brighter when a new component passes through it and slightly shifts its position when this happens, as we see in Fig.~\ref{BU_combined} (middle panel) and as predicted by numerical simulations \citep[e.g.,][]{Gomez1997}. Moreover, component \emph{d5} moves at an apparent velocity of 11.5$\pm$0.2$c$, in agreement with the apparent motions of the other detected moving components, and the range of speeds usually observed in this source \citep[e.g.,][]{Jorstad_b}. 

One of the 43 GHz epochs analyzed (19th January 2014) is only one day apart from the RA epoch presented in this work (18th January 2014). The total intensity image is shown in Fig.~\ref{43image}. At this epoch, the source displays a low-activity state, having a flux density peak at 43 GHz of only $\sim$1.58 Jy/beam. Also, component \emph{d5} shows a position compatible with component C from the 22 GHz RA image, visible in Fig. \ref{corezoom}.

Assuming that component \emph{c0} is the core, the fact that component \emph{d5} ejection epoch is subsequent to the \cite{Kovalev} one suggests that this should have been upstream of the core in January 2013. This, assuming a simple conical shape for the jet, would imply a size smaller than component \emph{c0} ($<40$ $\mu$as) thus similar to the size that \cite{Kovalev} attribute to the component producing the extreme brightness temperature (26 $\mu$as).
\cite{Marscher08} proposed that the mm-wavelength core of blazars could be a recollimation shock along the particle jet stream, while the region between the central black hole and the core would be responsible for the acceleration and collimation. In this context, we identify the \emph{c0} component as a recollimation shock, in addition to the one corresponding to the stationary component \emph{s4}. The two are separated by $\sim$0.2 mas. 

In Fig. \ref{BU_combined} (lower panel) we present the observed brightness temperature estimates for the different modelfit components versus time. The core (\emph{c0}) shows values between 10$^{11}$ and 10$^{12}$ K for most epochs. A similar trend is present for the stationary component \emph{s4}, while the newly-ejected one (\emph{d5}) starts from a $T_\mathrm{b}$ just below 10$^{11}$ K during the first epoch, and then progressively decreases with time. Nevertheless, the intrinsic brightness temperatures for each component/epoch, presented in Tab. \ref{modelfit}, result in values not exceeding the equipartition limit, in line with the estimates obtained from our RA observations at 22 GHz. The uncertainties in the flux density and distance from c0 of components have been calculated following the method described in \cite{CasadioCTA}, developed for VLBA observations at 43 GHz, and adopting a minimum error of 0.01 mas for the distance (half of beam size). The size uncertainty was calculated following \cite{Jorstad}.

\cite{Lisakov} presented a multi-epoch (2008-2014) VLBA study of 3C\,273 in the 5-43 GHz frequency range. The brightness temperature estimates therein were performed adopting a resolution limit for the modelfit components; this results in slightly higher $T_b$ than what we obtain with our approach at the same frequency. Also, the number of components used was different; in our case, we tried to minimize the number of free parameters (and thus components) when modeling the emission. This more conservative approach avoids a possible overestimate of the brightness temperature that could arise from arbitrarily small modelfit components. Nevertheless, the same $T_b$ trend is visible for the overlapping epochs (2013-2014), showing a maximum decrease of more than 2 orders of magnitude with respect to the \cite{Kovalev} epoch.



\subsection{Spectral index map}

\begin{figure}[]
\centering
\includegraphics[width=9cm]{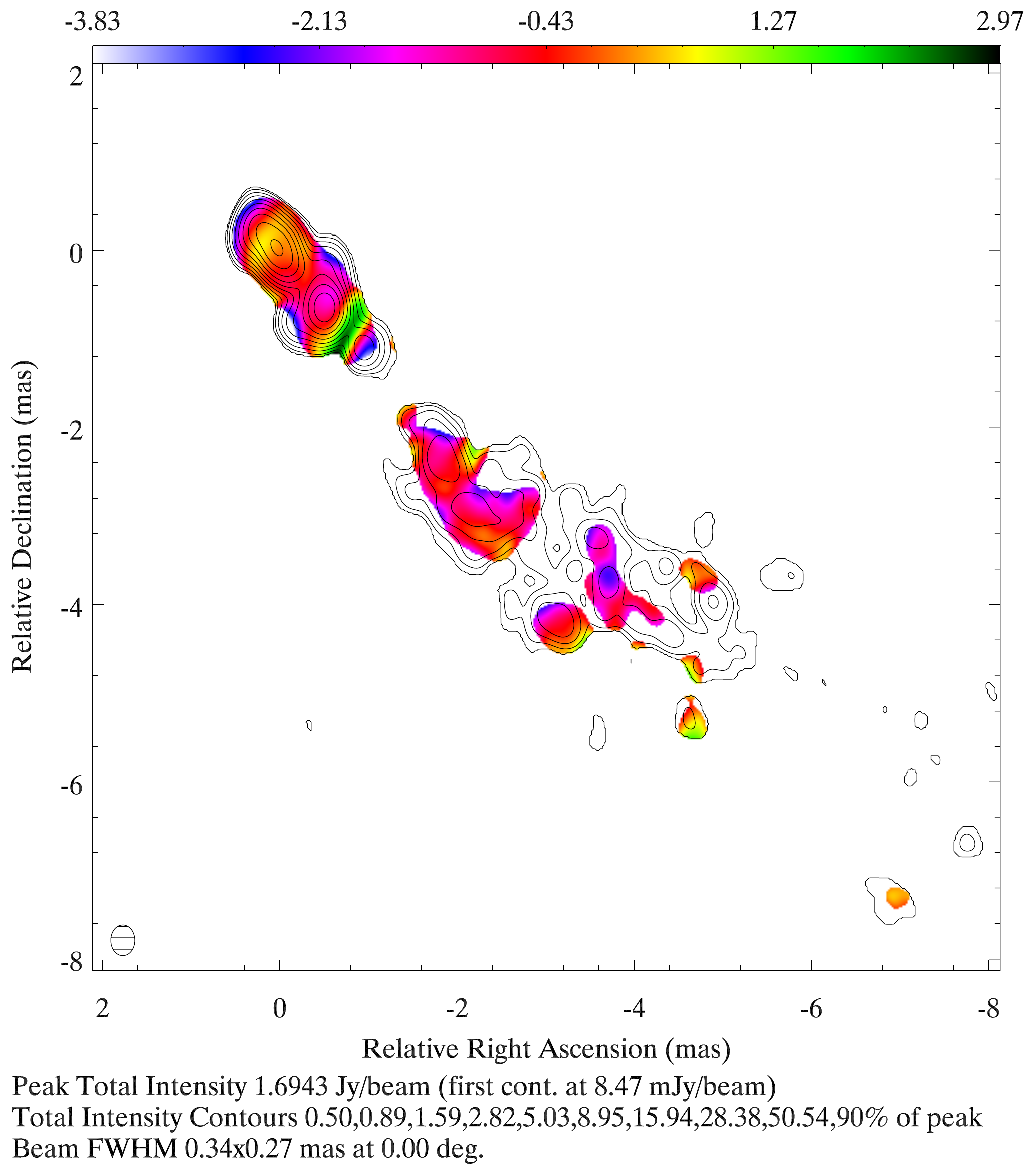}
\caption{Spectral index map obtained using the 22 GHz map from RA observations in this work, and the VLBA-BU-BLAZAR 43 GHz map at the epoch nearest in time (19th January 2014). Spectral index $\alpha$ is defined positive ($S\propto\nu^{\alpha}$).}
\label{spix}
\end{figure}

Fig. \ref{spix} shows the spectral index map between 22 and 43 GHz, realized using the map from the VLBA-BU-BLAZAR epoch only one day apart from our RA observations. Image registration was performed via cross-correlation analysis of the total intensity maps, as in \cite{gomez} (and references therein) applying a shift of [-0.12;0.12] mas in right ascension and declination, respectively. Spectral index $\alpha$ is defined as positive ($S\propto\nu^{\alpha}$).
A spectral index of $\alpha=0\pm0.1$, indicating a flat/optically-thick region, is visible within the \emph{c0} component. This is consistent with our interpretation of \emph{c0}  (component A in the RA map) as the core. The spectra in the regions corresponding to the position of \emph{s4} and \emph{d5} components (B and C components, respectively, in the RA map) have an $\alpha=0.3\pm0.1$ and $\alpha=-1.3\pm0.2$, respectively, indicating another optically thick region for \emph{s4} (the recollimation shock), and an optically thin region for \emph{d5} (the newly ejected component). Finally, the more extended emission towards southwest is steep as well, as expected in this frequency range for the more adiabatically expanded, optically thin, components of the jet.

\begin{figure*}
\begin{center}
\includegraphics[width=18cm]{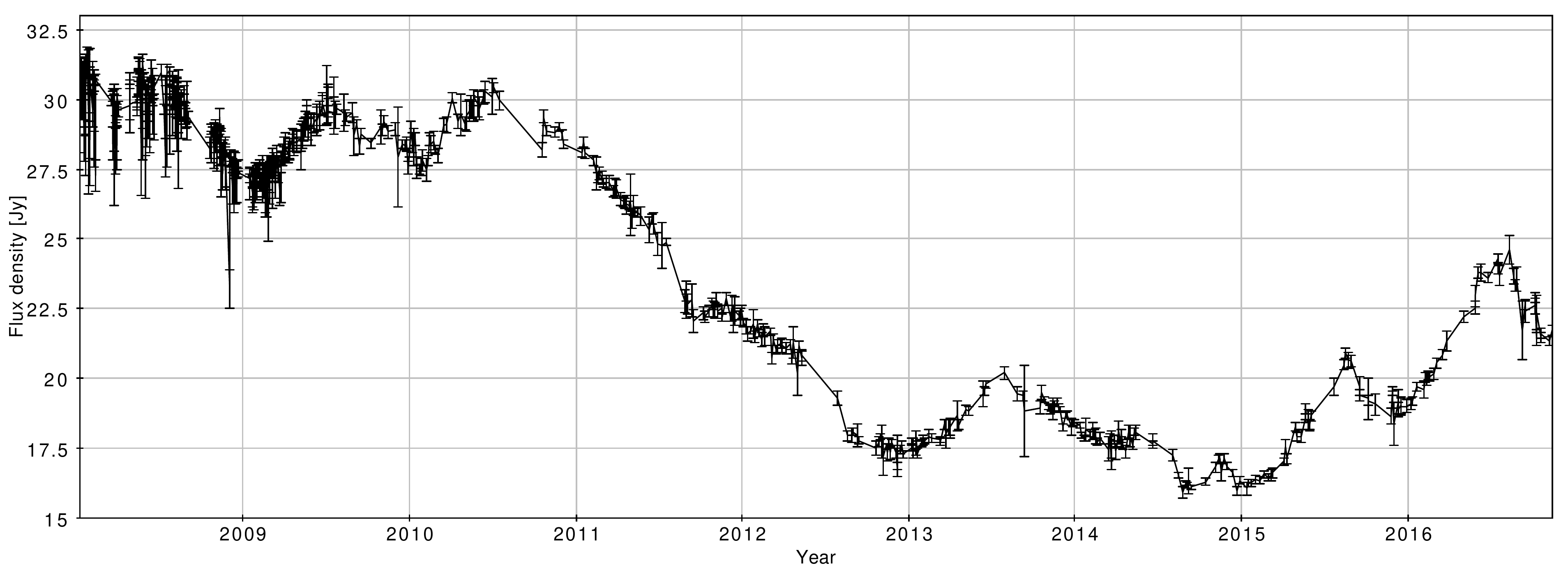}
\caption{Total flux densities from the OVRO 40-m single-dish monitoring program at 15 GHz, since 2008.}
\label{OVRO}
\end{center}
\end{figure*}


\section{Discussion}

In the following, we put our results in the context of previous RA observations and studies at different bands.

\subsection{Comparison with previous RadioAstron observations}

Drastic differences, in terms of brightness temperature estimates, are seen when comparing with previous RA observations of 3C\,273 at the same frequency, performed between December 2012 and February 2013 during the early-science program (\citealt{Kovalev}). An exceptionally high brightness temperature was detected - $\sim$10$^{13}$ K, related to a component of only 26 $\mu$as - challenging our comprehension of the jet physics. The space-baselines segment of these observations was conducted using the Arecibo, Green Bank, and VLA telescopes as the ground array. The highest brightness temperature measured at 22 GHz was 1.4$\times$10$^{13}$ K (rest- frame), on a baseline of 7.6 G$\lambda$. Assuming the same Doppler factor as for our observations, this gives an intrinsic brightness temperature of $\sim$10$^{12}$ K. Our observations, taken one year after, show that the brightness temperature has dropped by $\sim$2 orders of magnitudes. 

It is worth noting that no other RA experiment resulted in successful space-baseline fringes on 3C\,273, at 22 GHz, between the \cite{Kovalev} epoch and ours. In fact, the source has been regularly targeted by the RA AGN-survey KSP, spanning space-baselines between $\sim$4.5 and $\sim$25 G$\lambda$. This confirms the minimum-activity status of the source resulting from our observations.

These results support one of the scenarios proposed by \cite{Kovalev}, that is, that an exceptionally high brightness temperature, over the Compton catastrophe limit, can be caused by a transient phenomenon responsible for the departure from equipartition of energy between magnetic fields and particles. In particular, following \cite{Readhead}, the magnetic field energy density ($u_{B}$) varies as:
\begin{equation}
u_{B}=\frac{B^2}{8\pi}\propto T_{\mathrm{b,int}}^{-4},
\end{equation}
where $B$ is the magnetic field, and $T_\mathrm{b,int}$ is the intrinsic brightness temperature, while the particle energy density ($u_\mathrm{p}$) varies as:
\begin{equation}
u_{p}\propto T_{\mathrm{b,int}}^{4.5}.
\end{equation}
For values of $T_\mathrm{b}$ near or above the inverse Compton catastrophe $B$ has very low values - significantly increasing the synchrotron loss time - and particle energy density dominates the system, while for values below the equipartition ones, $u_B$ dominates. As a rough approximation, and assuming the peak frequency of the synchrotron emission spectrum did not change, 
for a $T_\mathrm{b,int}$ drop of two orders of magnitude between the two epochs considered here, the $B$ value must have increased by a factor $\sim10^4$ in order to restore the equipartition conditions. Although, given the presence of a newly ejected component between the two epochs, and the consequent probable change in particle energy density, this can only be a simple approximation.


\subsection{Source activity monitoring from OVRO single-dish data}
\label{sec:OVRO}

In Fig. \ref{OVRO} we report measurements at 15 GHz from the Owens Valley Radio Observatory (OVRO) monitoring project. Measurements from single-dish telescopes are particularly useful to study the total flux density variation of a given source, while at the resolution of VLBI part of the emission can be resolved out and not accounted for. The OVRO data span $\sim$8 years, starting from 2008. It is clearly visible how the total flux density of the source has dramatically dropped between 2011 and 2013, reaching an absolute minimum in August 2014 ($\sim$16 Jy), which is half of the flux density registered at the maximum ($\sim$32 Jy, January 2008). Since the RA observations presented in this work have been taken at a higher frequency (22 GHz), there could be a time-lag with respect to the minimum detected at 15 GHz, the drop at higher frequency being detected months before the one in the OVRO data.


\subsection{$\gamma$-ray emission detected by Fermi-LAT}

Since the launch of the {\emph{Fermi}} mission, carrying the Large Area Telescope (LAT), and the subsequent improvement of the $\gamma$-ray source census, several authors have found hints of correlation between the radio emission and $\gamma$-ray flares in Fanaroff-Riley I (FRI) and II (FRII) sources - sometimes linked to the ejection of a new superluminal component detected at mm-band (\citealt{Casadio, grandia, grandib}) -as well as in blazars (e.g., \citealt{CasadioCTA,Karamanavis,Schinzel}). For the blazar 3C\,273 in particular, \cite{Chidiac} found a significant correlation between flux variation in the $\gamma$-ray and mm bands, with an estimated time-lag of $\sim$110 days. \cite{Lisakov} found correlation between a $\gamma$-ray flare and the emission of a new component, allowing them to constrain the location of the $\gamma$-ray emission between 2 and 7 pc from the 7-mm core.

During our observations, the source was in a $\gamma$-ray-quiet state, spanning more than 3 years in the \emph{Fermi}-LAT data. The weekly light curve from the \emph{Fermi}-LAT collaboration\footnote{\url{http://fermi.gsfc.nasa.gov/ssc/data/access/lat/msl_lc/}} shows no flux measurements above 1$\times$10$^6$ photons/cm$^2$/s since $\sim$55500 MJD (November 2010), while a peak up to 5$\times$10$^6$ photons/cm$^2$/s was previously registered between 2009 and 2010. This seems to confirm the low-activity state of the source visible from radio data presented above (both VLBI and single-dish). Also, we notice that the passage of the newly-ejected component \emph{d5} in the mm-band core does not correlate with a flare in $\gamma$-ray band: \cite{Jorstad_c} reported that 62 out of the 114 superluminal knots identified in the VLBA-BU-BLAZAR observations from 2008 August to 2013 January were not associated with any significant gamma-ray flares, thus in agreement with what we observe. Furthermore, the lack of gamma-ray emission is consistent with the low $T_b$ measured.




\section{Conclusions}
We present 22-GHz RA observations of 3C\,273, obtained with the support of a global ground array of radio telescopes in the context of the AGN polarization KSP. We can summarize our conclusions as follows:

\begin{itemize}
 \item The source seems to be in a particularly low activity state, both from our observations and from OVRO single-dish monitoring at 15 GHz.
 \item We find a core brightness temperature at least two orders of magnitude lower than the one measured one year before by \cite{Kovalev} with RA SVLBI observations at the same frequency. 
 \item From the kinematics analysis of multi-epoch mm-band data, taken by the VLBA-BU-BLAZAR program, we detected a new component, ejected from the mm-core $\sim$2 months after the extreme brightness temperature found by \cite{Kovalev}. Thus, this was located upstream of the core during \cite{Kovalev} epoch, with an expected size $<$40 $\mu$as, compatible with the one estimated by those authors.
 \item No $\gamma$-ray flare seems to be present, either during 2013 or our epoch (2014). This is consistent with both the low measured brightness temperature of the knot and the finding by \cite{Jorstad_c} that fewer than 50\% of superluminal knots in blazars are associated with gamma-ray flares.
 \item In AGN jets, brightness temperatures above the equipartition limit, or even Compton-cooling limit, can be reached and maintained only
 for short periods. Our observation shows that a brightness temperature even lower than the equipartition value can be present after the extremely high value detected one year before. The access to extremely long baselines - and angular resolution - thus does not systematically find cm-band brightness temperatures larger than typical values found at lower resolution. This means that the source status still plays a principle role in the characterization of the observable physical quantities.
\end{itemize}


\begin{acknowledgements}

The \emph{RadioAstron} project is led by the Astro Space Center of the Lebedev Physical Institute of the Russian Academy of Sciences and the Lavochkin Scientific and Production Association under a contract with the Russian Federal Space Agency, in collaboration with partner organizations in Russia and other countries.

This paper includes data observed with the 100\,m Effelsberg radio-telescope, which is operated by the Max-Planck-Institut f\"ur Radioastronomie in Bonn (Germany)

The VLBA is an instrument of the Long Baseline Observatory.  The Long Baseline Observatory  is  a  facility  of  the  National  Science Foundation operated under cooperative agreement by Associated Universities, Inc.

The European VLBI Network is a joint facility of independent European, African, Asian, and North American radio astronomy institutes. Scientific results from data presented in this publication are derived from the EVN project code GA030C (ground array), and from the \emph{RadioAstron} mission project code RAKS04C (space segment).

JLG and AF acknowledge support by the Spanish Ministry of Economy and Competitiveness grants AYA2013-40825-P and AYA2016-80889-P.
AA and MAPT acknowledge support by the Spanish MINECO through grants AYA2012-38491-C02-02 and AYA2015-63939-C2-1-P, cofunded with FEDER funds.  ER acknowledges support from the Spanish MINECO through grants AYA2012-38491-C02-01 and AYA2015-63939-C2-2-P and from the Generalitat Valenciana grant PROMETEOII/2014/057.

TS was funded by the Academy of Finland projects 274477 and 284495.

The research at Boston University was supported in part by NASA Fermi Guest Investigator grant NNX14AQ58G.

SSL was supported by the National Research Foundation of Korea (NRF) grant funded by the Korea government (MSIP) (No. NRF-2016R1C1B2006697).

YYK was supported in part by the Alexander von Humboldt Foundation.

This research has made use of data from the OVRO 40-m monitoring program (Richards et al. 2011) which is supported in part by NASA grants NNX08AW31G, NNX11A043G, and NNX14AQ89G and NSF grants AST-0808050 and AST-1109911.

\end{acknowledgements}


\end{document}